# Robust Containerization of the High Angular Resolution Functional Imaging (HARFI) Pipeline

Zhiyuan Li[1], Kurt G. Schilling[2], and Bennett A. Landman[1,2,3]

[1] Department of Electrical and Computer Engineering, Vanderbilt University, Nashville, TN, USA
[2] Department of Radiology & Radiological Sciences, Vanderbilt University Medical Center, Nashville, TN, USA
[3] Department of Computer Science, Vanderbilt University, Nashville, TN, USA

## 1 Introduction

Historically, functional magnetic resonance imaging (fMRI) of the brain has focused primarily on gray matter, particularly the cortical gray matter and associated nuclei[1]. However, recent work has demonstrated that functional activity in white matter (WM) also plays a meaningful role in both cognition and learning. For example, task-based blood-oxygen-level dependent (BOLD) signal changes have been reported in the corpus callosum[2], internal capsule[3], and optic radiations[4]. Since these early observations, investigations into the relationship between functional activity in white matter and cognitive or neurological states—especially in task-based paradigms—have become an increasingly important area of research, with both biological and clinical implications. Statistical analysis of WM resting-state signals has revealed functional networks aligned with known WM pathways[5]. Notably, WM resting-state fMRI signals exhibit anisotropic directional neighborhood correlations that are, of particular interest here, largely consistent with WM bundle orientations derived from diffusion MRI and persist over long distances[6]. One particularly intriguing approach, functional-correlation tensor (FCT)[7], involves characterizing the local correlation structure of functional signals in white matter using tensor-based or constrained spherical harmonic representations. This local correlation structure parallels the architectural insights derived from diffusion tensor imaging (DTI) or high angular resolution diffusion imaging (HARDI). In previous work[8], we introduced the High Angular Resolution Functional Imaging (HARFI) pipeline, which demonstrated both local and global patterns of functional correlation in white matter. Notably, HARFI enabled exploration of asymmetric voxel-wise correlation using both even- and odd-order spherical harmonics and construction of functional orientation distributions (FODs). Although the original implementation of HARFI was released via GitHub, adoption was limited due to the technical complexity of running the source code. In this work, we present a robust and efficient containerized version of the HARFI pipeline, enabling seamless execution across multiple public datasets.

---

[1] John C Gore, "Principles and Practice of Functional MRI of the Human Brain," *The Journal of Clinical Investigation* 112, no. 1 (2003): 4–9; Massimo Filippi, *FMRI Techniques and Protocols*, vol. 830 (Springer, 2016).
[2] Ryan C N D'Arcy et al., "Exploratory Data Analysis Reveals Visuovisual Interhemispheric Transfer in Functional Magnetic Resonance Imaging," *Magnetic Resonance in Medicine: An Official Journal of the International Society for Magnetic Resonance in Medicine* 55, no. 4 (2006): 952–58.
[3] Jodie R Gawryluk et al., "Investigation of FMRI Activation in the Internal Capsule," *BMC Neuroscience* 12 (2011): 1–7.
[4] Lauren Marussich et al., "Mapping White-Matter Functional Organization at Rest and during Naturalistic Visual Perception," *Neuroimage* 146 (2017): 1128–41.
[5] Aviv Mezer et al., "Cluster Analysis of Resting-State FMRI Time Series," *Neuroimage* 45, no. 4 (2009): 1117–25; Michael Peer et al., "Evidence for Functional Networks within the Human Brain's White Matter," *Journal of Neuroscience* 37, no. 27 (2017): 6394–6407.

Our goal is to facilitate broader and deeper exploration of functional white matter architecture—especially through the lens of high angular resolution functional correlations. The key innovation of this work is the containerized implementation of HARFI, which addresses several usability challenges in the original GitHub code. We have released the new tool under a permissive open-source license to promote reproducible and accessible research. This implementation enables users to extract complex FODs from fMRI data and supports downstream analyses such as tractography.

## 2      Source Code Containerization

We began by investigating the original HARFI GitHub source code using a private dataset with precomputed HARFI outputs, kindly provided by the authors. During this reproduction effort, we encountered two challenges when running the GitHub code. The first challenge is a mismatch in the unit of the radius variable named R. In the HARFI paper, the unit of R (referred to as $R_{paper}$ in the following text) is millimeters (mm), whereas in the GitHub code, the unit of the parameter R (referred to as $R_{code}$) is in voxels. Therefore, to use the recommended radius value from the HARFI paper ($R_{paper}$ = 9 mm), it must be converted to voxel units when running the HARFI GitHub code. For example, one should set $R_{code}$ to 3 voxels for an fMRI image with 3 mm isotropic resolution to match the target $R_{paper}$ value of 9 mm. The second challenge we encountered was that a section of the code responsible for the detrending process had been accidentally commented out in a previous commit of the HARFI GitHub repository. Based on the description in the HARFI paper and confirmation from the authors, we restored this missing code segment. After addressing these two issues, we were able to successfully reproduce the exact outputs provided by the authors.

---

[6] Zhaohua Ding et al., "Spatio-Temporal Correlation Tensors Reveal Functional Structure in Human Brain," *PloS One* 8, no. 12 (2013): e82107.

[7] Zhaohua Ding et al., "Visualizing Functional Pathways in the Human Brain Using Correlation Tensors and Magnetic Resonance Imaging," *Magnetic Resonance Imaging* 34, no. 1 (2016): 8–17.

[8] Kurt G Schilling et al., "Functional Tractography of White Matter by High Angular Resolution Functional-Correlation Imaging (HARFI)," *Magnetic Resonance in Medicine* 81, no. 3 (2019): 2011–24.

With the corrected source codes, we then aimed to build a containerized implementation for the HARFI pipeline. The corrected HARFI pipeline was developed and tested in MATLAB application (MathWorks, Natick, MA, USA) using version R2022b Update 10, on Ubuntu 22.04 LTS. To ensure ease of execution and usability, we compiled the MATLAB source code into a standalone executable using MATLAB compiler (mcc), which can be run with the MATLAB Runtime (MCR). Subsequently, we packaged the MATLAB executable within a Singularity[9] container that includes all required environments and libraries. This makes the containerized HARFI tool conveniently accessible without requiring users to manage MATLAB installation or configure additional dependencies. Figure 1 illustrates the comprehensive workflow and key functionalities of the HARFI pipeline in the containerized implementation. The HARFI pipeline transforms resting-state fMRI data into voxel-wise FODs derived from BOLD signal directional correlations. The entire process can be broken down into the following sequential steps.

## 2.1 Preprocessing (User-Provided)

The HARFI pipeline assumes that users have already preprocessed their 4D fMRI volumes prior to running HARFI[10]. Slice-timing correction and motion correction are recommended as common preprocessing steps of fMRI data. Users are encouraged to utilize established preprocessing tools such as SPM12 or FSL to perform these tasks. We leave the freedom to allow users the flexibility to apply their preferred preprocessing methods and parameters, and these steps are not implemented within the containerized HARFI implementation. A corresponding brain mask, registered to the fMRI space, is required to run HARFI pipeline.

## 2.2 Detrending and filtering

The pipeline takes the input preprocessed fMRI data and then the linear trends from the BOLD signals are removed to correct signal drifts. Additionally, temporal band-pass filtering is performed to retain frequencies between 0.015 Hz and 0.1 Hz. The low- and high-pass filter are both implemented via Chebyshev Type II filter in MATLAB. To avoid problems introduced by magnetic equilibration transients, the first N frames of each fMRI data (typically N = 20 as we recommend) are discarded.

---

[9] Gregory M Kurtzer, Vanessa Sochat, and Michael W Bauer, "Singularity: Scientific Containers for Mobility of Compute," *PloS One* 12, no. 5 (2017): e0177459.
[10] Schilling et al., "Functional Tractography of White Matter by High Angular Resolution Functional-Correlation Imaging (HARFI)."

### 2.3 Directional Sampling

HARFI computes the orientational correlation of BOLD signals by first sampling the local neighborhood of each voxel along certain directions on a sphere. In the proposed pipeline, a set of 60 uniformly distributed unit vectors is defined, and is provided in Jones60.mat in the Singularity container. A maximum radius in voxel $R_{code}$ is chosen for a target of 9 mm for direction integration. For example, one should set $R_{code}$ to 3 for fMRI data with 3 mm isotropic resolution. A target of 8 mm or 10 mm also works well. Our implementation then uses linear interpolation of the fMRI signals to sample intermediate points in each given direction.

### 2.4 Directional Integration

After directional sampling, for each voxel in the mask, correlation coefficients are calculated for its neighborhood voxels in the distance up to $R_{code}$ along each direction. Then integration over distance is calculated to provide the final single-value measure in each direction. The result of this stage is a 4D array representing, for every voxel in the mask, the integrated correlation along each of the 60 pre-defined directions on the sphere.

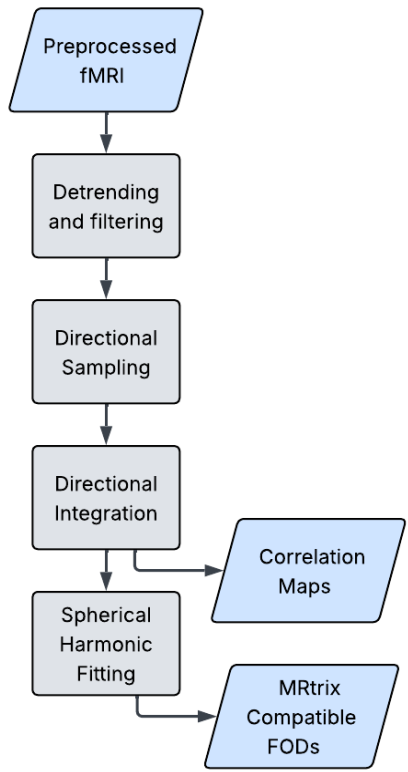

Figure 1. The workflow and functionalities in the HARFI containerized implementation. Given the input fMRI data, HARFI calculates and saves the functional correlation maps and its corresponding functional orientation distributions.

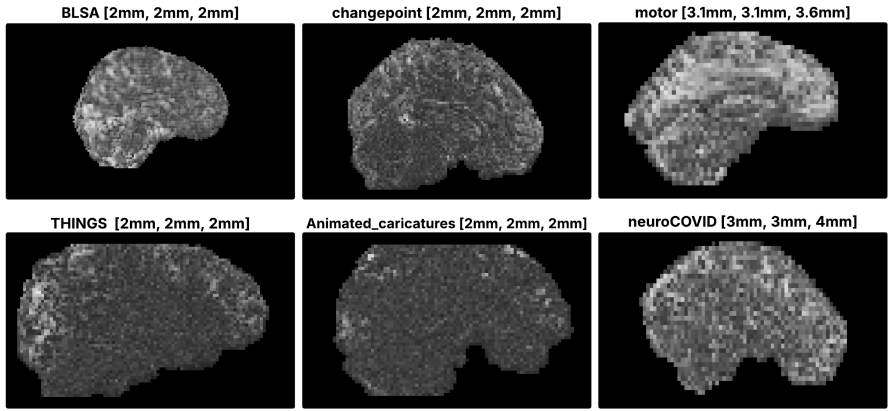

Figure 2. Functional correlation maps are calculated by the proposed HARFI Singularity container for six fMRI scans. Each scan's spatial resolution (denoted in brackets) and field of view (FOV) are unmodified for visualization. Bright voxels denote high correlation, while dark voxels denote low correlation. Different brain activities exhibit varying correlations among brain regions.

## 2.5 Spherical Harmonic Fitting

We next fit spherical harmonic (SH) models to each voxel's correlation map considering all the 60 directions. Two sets of SH coefficients are computed, which are (1) Even-order-only fit that enforces antipodal symmetry, and (2) Full (even + odd) order fit that aims to capture potential asymmetry in functional correlations. This stage is implemented using an internal MATLAB SH toolbox, and the SH coefficients are solved via regularized least-squares fitting. The result is a compact and smooth representation of the voxel-wise FODs, capturing principal functional connectivity orientations at each voxel.

## 2.6 Export to MRtrix Format and Downstream Analysis (User-Defined)

The fitted SH coefficients are reformatted into MRtrix-compatible FOD volumes by adjusting the order of the SH coefficients to meet MRtrix's standard. These can then be used directly as input to tractography tools such as tckgen of MRtrix3, to produce functional pathways. While the proposed HARFI pipeline produce FODs ready for tractography, additional quantitative analyses are not included in the Singularity. Users may develop metrics and analyze the fitted SH coefficients at their own interests.

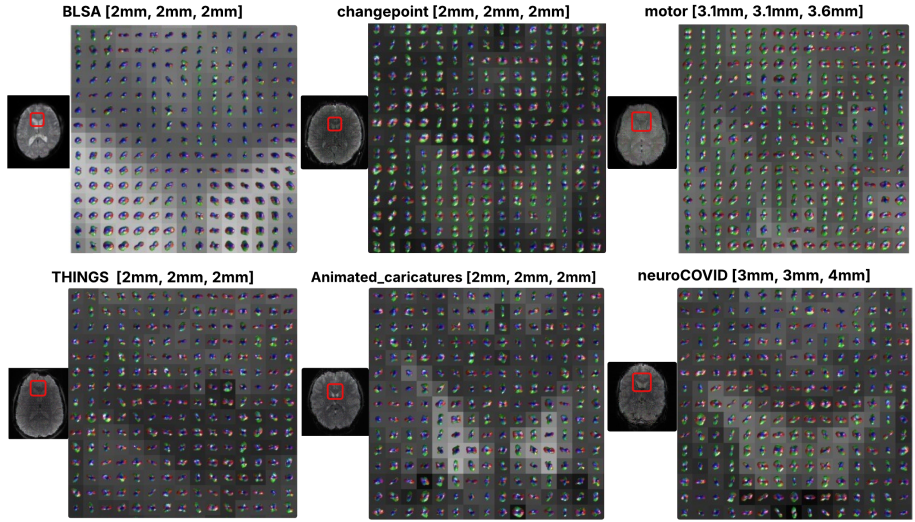

Figure 3. Functional orientation distributions are calculated by the proposed HARFI Singularity container. HARFI fits are shown for an axial slice from each of the six scans, with magnified views highlighting crossing or complex distributions.

## 2.7 Singularity Container

Finally, a Singularity[11] container was built for the proposed pipeline using version 3.8.1, and it was tested with both Singularity versions 3.8.1 and 4.3.1 on Ubuntu 22.04 LTS. First, we wrote a new MATLAB function that integrates all pipeline procedures with updated and corrected source code. This function was then compiled into a standalone executable using the MATLAB Compiler (mcc) in MATLAB R2022b Update 10. Subsequently, the executable was encapsulated within a Singularity container, and a "runscript" was created to invoke the executable automatically. In other words, the complete HARFI pipeline is executed automatically whenever users run the container using the "singularity run" command. Additionally, the MATLAB Runtime (MCR) libraries for MATLAB R2022b Update 10 were installed in the container to ensure proper execution of the compiled code. Other source files were also included in the Singularity container for users' reference.

Four key parameters of HARFI can be adjusted in the input: (1) N – the number of initial fMRI frames to discard, (2) TR – Repetition Time (in seconds), (3) Discrete – the number of discrete steps in integration, and (4) R – the radius of integration in voxels.

---

[11] Kurtzer, Sochat, and Bauer, "Singularity: Scientific Containers for Mobility of Compute."

## 3      Experiments

To validate the proposed HARFI Singularity container, we tested it on the BLSA dataset[12] and five public datasets on OpenNeuro, which are Animated caricatures fMRI study[13], THINGS-fMRI[14], motor-fmri[15], neuroCOVID-fMRI[16], and changepoint-fMRI[17]. Following the validation methods described in the HARFI paper, we qualitatively present visualizations of the correlation maps and the FODs generated by the HARFI fitting. Expected HARFI outputs should exhibit diverse correlation values among brain regions and capture complex FODs, as demonstrated by both single-directional and crossing glyphs in the SH visualizations. Fig. 2 shows that the proposed HARFI Singularity container can successfully calculate functional correlation maps across fMRI scans and reflect various patterns of functional correlation associated with different brain activities. Fig. 3 demonstrates that the proposed HARFI Singularity container can successfully capture complex, asymmetric functional distributions.

## 4      Information Sharing Statement

The presented tool is publicly available as open source under the Creative Commons License at https://github.com/MASILab/HARFI_Singularity.

---

[12] Nathan Wetherill Shock, *Normal Human Aging: The Baltimore Longitudinal Study of Aging* (US Department of Health and Human Services, Public Health Service, National Institutes of Health, National Institute on Aging, Gerontology Research Center, 1984).

[13] N Furl et al., "'Animated Caricatures FMRI Study'" (OpenNeuro, 2020), https://doi.org/10.18112/openneuro.ds002741.v1.0.2.

[14] Martin N Hebart et al., "'THINGS-FMRI'" (OpenNeuro, 2024), https://doi.org/doi:10.18112/openneuro.ds004192.v1.0.7.

[15] J Veillette and H Nusbaum, "'motor-Fmri'" (OpenNeuro, 2024), https://doi.org/doi:10.18112/openneuro.ds005239.v1.0.1.

[16] Alejandra Figueroa-Vargas et al., ""neuroCOVID MRI DWI and FMRI with Reversal Learning "" (OpenNeuro, 2024), https://doi.org/doi:10.18112/openneuro.ds005364.v1.0.0.

[17] Chang-Hao Kao et al., "'changepoint Fmri'" (OpenNeuro, 2020), https://doi.org/10.18112/openneuro.ds003170.v2.0.0.